\documentstyle[pre,preprint,aps]{revtex}
 
\newcommand{\displayfrac}[2]{\frac{\displaystyle #1}{\displaystyle #2}}
 
\makeatletter
\newcommand{\newoperator}[2]{\@ifdefinable#1{\def
 #1{\mathop{#2}\nolimits}}}
\newcommand{\newoperatorwithlimits}[2]{%
\@ifdefinable#1{\def#1{mathop{#2}}}}
\makeatother
 
\newoperator{\diff}{\rm\mathstrut d\!}

\begin{document}
\draft
 
\title{Developments in Time Analysis of Particle and Photon \\
Tunnelling}
 
\author{V. S. Olkhovsky}
\address
{\it Institute for Nuclear Research, National Academy of Sciences \\
of the Ukraine, 252028 Kiev, Ukraine}
\author{A. Agresti}
\address{\it Dipartimento di Fisica, Sezione di
Fisica Superiore dell'Universit\`a
di Firenze,\\ via S. Marta 3, 50139 Firenze, Italy}

\maketitle
\narrowtext
 
\begin{abstract}
A new systematization of various theoretical approaches to defining
tunnelling times for nonrelativistic particles in the light of time as a
quantum--mechanical observable is given. Then new results on the analogy
between particle and photon tunnelling and on time as an observable in
quantum electrodynamics and also analysis of the causality validity
during tunneling are presented.
\end{abstract}

\vspace{1 cm}
\section{Introduction}
Tunnelling time analysis has a long history. The problem of defining the
tunnelling time was posed already in the beginning of the 30--th
\cite {con,mac}.
But from then it was remained almost ignored until the 50--60--th when
the more general problem of defining the quantum--collision duration
began to be investigated
\cite{boh,wig,smi60,gold,ohm,fon,baz,ryb,olk67,jau67,ved,olkrec69}
 and almost simultaneously, after a
long period of the silence from the 20--th induced by the Pauli theorem
about the impossibility to construct the self--adjoint time operator in
quantum mechanics \cite{pau}, the first attempts appeared to introduce
the notion of time as a quantum--mechanical observable
 \cite{eng59,pau62,lip,raz,gie,ros,all}. And
during the 70--80--th (mainly in \cite{olk74,hol,olk84}) the membership
of time in
the set of quantum--mechanical observables had been principally cleared
up. The detailed analysis of developments in the study of time as a
quantum--physical observable is contained in \cite{olk84,olk90,agru}.
 
The developments in physics of condensed media, physics of
electromagnetic--wave propagation, biophysics and especially the advent
of high--speed electronic devices, based on tunnelling processes,
revived an interest in the tunnelling time analysis, whose relevance has
always been apparent in physics of nuclear sub--barrier fission and
fusion, and stimulated the publication of not only a lot of theoretical
studies but already a number of theoretical reviews
\cite{col89,hau89,hua90,but90,lea90,jau91,jon,olkrec92,land94}.
 
Regarding the experimental research of tunnelling times, the great
difficulties of real measurements for particles (in particular, too
small values of tunnelling times) made the verifications of theoretical
results to be practically impossible (see, for instance,
\cite{ngu,ran91}).
Only recently there were realized some measurements of tunnelling times
for microwaves and laser--light photons \cite{end92,ste93,ran93,spi94}.
In what sense
the results of these experiments are suitable for the time analysis of
particle tunnelling ? Although the formal analogy of particle and photon
(electromagnetic wave--packet) tunnelling is well seen by simple
comparing the relevant stationary equations
\cite{chi91,ran92,mar92,nim94}, actually we
deal with the {\it time--dependent} equations and moreover, the problem
of time as an observable also in quantum electrodynamics must be
resolved too. Below (in Sec.7) these questions will be explicitly
analyzed.
 
Returning to the problem of the theoretical definition of the tunnelling
time for particles, we see not only the absence of the consensus in such
definition but also declarations about the incompatibility of some
approaches both quantitatively and in the physical interpretation
\cite{col89,hau89,hua90,but90,lea90,jau91,jon,olkrec92,land94}. Among
the reasons of such situation there are the
following ones: {\it (i)} the problem of defining the tunnelling time is
closely connected with general fundamental problems of time as a {\it
quantum--physical observable} and the general definition of
quantum--collision durations. And the acquaintance with the principal
solution of these problems have not got a wide prevalence yet. {\it
(ii)} The motion of particles inside a potential barrier is a quantum
phenomenon {\it without} any direct classical limit. {\it (iii)} There
are essential {\it differences} in {\it initial, boundary} and
{\it external conditions} of various definition schemes which have not
been systematically analyzed yet.
 
Following \cite{land94,pri}, we can arrange the majority of approaches
into
several groups which are based on: (1) the time--dependent wave--packet
description; (2) averaging over an introduced set of kinematics paths,
distribution of which is supposed to describe the particle motion inside
the barrier; (3) introducing a new degree of freedom, constituting a
physical clock for measurements of tunnelling times. Separately, by
one's self, the dwell time stands. The last has {\it ab initio} the
presumptive meaning of the time that the incident flux has to be turned
on, to provide the accumulated particle storage in the barrier
\cite{smi60,land94}.
 
The first group contains the so--called phase times, firstly mentioned
in \cite{boh,wig} and applied to tunnelling in \cite{har,fle}, the times
of the motion of wave--packet spatial centroids, earlier considered for
general quantum collisions in \cite{ved,olkrec69} and applied to
tunnelling
in \cite{hau87,lan92}, and finally the Olkhovsky--Recami (O--R) method
\cite{olkrec92,olkrec94,olkrec95} of averaging over
unidirectional fluxes, basing on the
representation of time as a quantum--mechanical observable and on the
generalization of the definitions,introduced in
\cite{ohm,olk84,olk90}
for atomic and nuclear collisions. The second group contains methods,
utilizing the Feynman path integrals
\cite{sok87,pol84,sok90,mug88}, the Wigner
distribution paths \cite{lee,muga92} and the Bohm approach
\cite{leav93}. To
the third group the approaches with the Larmor clock \cite{butt83} and
the oscillatory barrier \cite{butt82,pim} pertain.
 
Certainly the basic self--consistent definition of tunnelling durations
(mean values, variances of distributions and so on) has to be elaborated
quite {\it similarly} to the definitions of other physical quantities
(distances, energies, momenta, etc.,) on the base of utilizing all
necessary
properties of time as a quantum--physical observable (time operator,
canonically conjugated to energy operator; the equivalency of the
averaged quantities in time and energy representations with adequate
measures, or weights, of averaging). Since the representation of
solutions of the time--dependent Schr\"odinger equation as moving wave
packets is typical and the most self--consistent in quantum collision
theory  (see, for instance, \cite[ref.3]{gold}, it is natural
to apply the wave--packet description. Then one can hope that in the
framework of the conventional quantum mechanics every known definition
of tunnelling times can be shown, after appropriate analysis, to be (at
least in the asymptotic
region, used for typical boundary conditions in quantum collision
theory) either a {\it particular} case of the general definition or an
{\it equivalent} one or the definition which is valid {\it not} for
tunnelling but for some accompanying process, different from tunnelling.
 
Here such a definition with the necessary formalism is presented (Sec.2)
 and, without claiming to present the exhaustive analysis of all known
definitions, then analysis of various approaches is given (Sec.3--5),
basing on the O--R formalism. In Sec.6 some peculiarities of tunnelling
evolution revealed by use of the O--R method are presented, then, after
Sec.7, there is a short note on the reshaping (and reconstructing)
phenomenon in connection with a possible formulation of relativistic
causality in the cases of superluminal effective tunnelling velocities
(Sec.8) and finally in Sec.9 some conclusions and reasonings on
nearest prospects are presented.
 
\section{The Olkhovsky--Recami formalism of defining tunnelling
durations, based on utilizing properties of time as a
quantum--mechanical observable and the wave--packet description.}
 
We confine ourselves to the simplest case of particles moving only along
the $x$--direction, and consider a time--independent barrier in
the
interval $(0, a)$; see Fig. 1, in which a larger interval
$(x_{i}
, x_{f})$, containing the barrier region, is also indicated. Following
the definition of collision durations, put forth in
\cite{ohm,olk74,olk84,olk90} and generalized in
\cite{olkrec92,olkrec94,sok87}, we can eventually define the mean values
of the time at which a particle passes through position $x$,
travelling in the positive or negative direction of the $x$--axis,
and the variances of the distributions of these times, respectively, as
\begin{equation}
< t_\pm (x) > \  = \  {{\displaystyle\int_{-\infty}^{+\infty} t
J_\pm (x,t)
\diff x} \over  {\displaystyle\int_{-\infty}^{+\infty} \ J_{\pm}
(x, t) \diff x}}
\label{1}
\end{equation}
and
\begin{equation}
{\rm D} \, t_\pm (x)  \  = \
{{\displaystyle\int_{-\infty}^{+\infty} t^2 J_\pm (x,t)
\diff x}
\over  {\displaystyle\int_{-\infty}^{+\infty}  \ \ J_{\pm}
(x, t) \diff x}} - [ < t_\pm (x) > ]^2
\label{2}
\end{equation}
where $ J_\pm (x,t) = {\rm Re} [(i \hbar / m)( \Psi (x,t) ) ( \partial
\Psi^*
 (x,t) / \partial x)] $ being the probability flux density for an
evolving wave packet $\Psi (x,t)$. We recall here the equivalence of
time and energy representations, with appropriate measures of averaging,
in the following sense: $< \ldots >_t = < \ldots >_E$  (index $t$ is
omitted in all expression for $< \ldots >_t$ for the sake of the
simplicity). This equivalence is a consequence of the time--operator
existence. So, we have a formalism for defining mean values, variances
and other central moments related to the duration distributions of all
possible kinds of collisions and interactions with arbitrary energies,
including tunnelling. For instance, for transmissions from region $I$
 to region $III$, we have
\begin{equation}
< \tau_T (x_i , x_f ) > = < t_+ ( x_f ) > - < t_+ ( x_i ) >
\label{3}
\end{equation}
\begin{equation}
{\rm D} \tau_T (x_i , x_f )  = {\rm D} t_+ ( x_f )  - {\rm D} t_+ ( x_i
)
\label{4}
\end{equation}
with $-\infty < x_i \le 0$ and $a \le x_f < \infty$. For a pure
tunnelling process one has
\begin{equation}
< \tau_{tun} (0, a) > = < t_+ ( a ) >  - <  t_+ (0) >
\label{5}
\end{equation}
and
\begin{equation}
{\rm D} \tau_{tun} (0, a )  = {\rm D} t_+ ( a )  + {\rm D} t_+ (
0 ).
\label{6}
\end{equation}
 
Similar expression we have for the penetration (into the barrier region
$II$) temporal quantities $ < \tau_{pen} (x_i, x_f ) > $ and $ {\rm D}
\tau_{pen} (x_i, x_f ) $ with $ 0< x_f < a$. For reflections in any
point $ x_f  \le  a$ one has
\begin{equation}
< \tau_R (x_i , x_f ) > = < t_- ( x_f ) > - < t_+ ( x_i ) >
\label{7}
\end{equation}
and
\begin{equation}
{\rm D} \tau_R (x_i , x_f )  = {\rm D} t_- ( x_f )  + {\rm D} t_+ ( x_i
).
\label{8}
\end{equation}
 
We stress that these definitions hold within the framework of
conventional quantum mechanics, {\it without} introducing any new
physical postulate.
 
In the asymptotic cases, when $|x_i| >> a $ ,
\begin{equation}
< \tau_{T}^{as} (x_i , x_f ) > = < t ( x_f ) >_T  - < t ( x_i ) >_{in}
\label{3a}
\end{equation}
and
\begin{equation}
< \tau_{T}^{as} (x_i , x_f ) > = < \tau_T ( x_i, x_f ) >  + < t_+  ( x_i
) >  - < t_+ (x_i) >_{in}
\label{9}
\end{equation}
where $ < \ldots >_T$ and $ < \ldots >_{in}$ denote averagings over the
fluxes corresponding to $ \psi_T  = A_T {\rm exp} (ikx)$
and $\psi_{in} = {\rm exp} (ikx) $ respectively.
 
For initial wave packets
\begin{eqnarray*}
\Psi_{in} (x, t) = \int_{0}^{\infty} G (k - \overline k) \exp [ikx -
i Et / \hbar ) {\diff E}
\end{eqnarray*}
(where $ E= \hbar^2 k^2 / 2m; \int_0^{\infty} |G (k- \overline k ) |^2
 {\diff E} = 1; G(0)=G(\infty)=0; k>0 $)
with sufficiently small energy (momentum) spreads, when
\begin{eqnarray*}
\int_0^\infty v^n |G A_T |^2 {\diff E} \cong  \int_0^\infty  v^n
|G|^2 {\diff E}  \ \ \ \ \ \  n= 0, 1,
\end{eqnarray*}
we get
\begin{equation}
< \tau_{T}^{as} (x_i , x_f ) > \cong < \tau_T^{ph} ( x_i, x_f ) >_E
\label{10}
\end{equation}
where
\begin{eqnarray*}
< \ldots  >_E \  = \  {{\displaystyle\int_0^{\infty}
v |G(k- \overline k)|^2
\diff E} \;\; \{ \ldots \} \; \over  {\displaystyle\int_0^{\infty} \
v |G (k- \overline k) |^2  \diff E}}
\end{eqnarray*}
and
\begin{equation}
\tau_T^{ph} (x_i, x_f) = (1/v) (x_f - x_i ) + \hbar
d ({\rm arg}A_T) / dE
\label{11}
\end{equation}
is the phase transmission time obtained by the stationary--phase
approximation. At the approximation when (\ref{10}) is valid and with a
small contribution
of $Dt_+ (x_i)$ into the variance $D \tau_T (x_i, x_f)$ (that can be
realized for sufficiently large energy spreads, i.e., short wave
packets), we get
\begin{equation}
D \tau_T (x_i, x_f) = \hbar^2 < (d |A_T| / dE)^2 >_E  \Bigr/ < |A_T|^2
>_E. \label{12}
\end{equation}
For the opposite case of very small energy spreads (quasi--monochromatic
particles) expression (\ref{12}) becomes the
item of $Dt_+ (x_f)$ and $D \tau_T (x_i, x_f)$ which is born by the
barrier influence.
 
When $|G|^2 \rightarrow \delta (E - {\overline E}), {\overline
E}$ being $ \hbar^2 {\overline k}^{\; 2} / 2m$, we get for
$ < \tau_T^{as} (x_i, x_f ) >  \cong <~\tau_T^{ph} (x_i, x_f) >_E$
strictly the ordinary phase time, without averaging. For a rectangular
barrier with height $V_0$ and $ \chi a >> 1$ ( where $ \chi =  [2m ( V_0
- E )]^{1/2}  /  \hbar)$, the expressions ( \ref{10} ) and ( \ref{12} )
 for $ x_i=0$ and $x_f= a$ pass, in the same limit, into the known
expressions
\begin{equation}
\tau_{tun}^{ph}  =  2 / v  \chi
\label{11a}
\end{equation}
( see \cite{har} and also \cite{olkrec92,land94} )  and
\begin{equation}
( D \tau_{tun}^{ph} )^{1/2} = a k / v \chi
\label{12a}
\end{equation}
(coincident with one of the Larmor times \cite{butt83} and the
B\"{u}ttiker--Landauer
 time \cite{pim} and also the imaginary part of the
complex time in the Feynman path--integration approach) respectively
(see also \cite{huang88}).
 
For the {\it real} weight amplitude $G(k- {\overline k})$, when
$< t(0)>_{in} = 0 $, from (\ref{9}), we obtain
\begin{equation}
< \tau_{tun} (0, a) > = < t_{tun}^{ph}  >  - <  t_+ (0) >.
\label{13}
\end{equation}
By the way, if the measurement conditions are chosen to be such that
only the positive-- momentum components of wave packets are registered,
i.e., \begin{math}
\Lambda_{exp,+} \Psi (x_i, t) =  \Psi_{in} (x_i, t) ;
\Lambda_{exp, +} \end{math} being
the projector onto positive--momentum states, then for any $x_i$ from
$(-\infty , 0 )$ and $x_f$ from $(a, \infty )$
\begin{equation}
< \tau_T (x_i , x_f ) >_{exp} = < t^{ph}_T ( x_i, x_f ) >_E
\label{3b}
\end{equation}
and
\begin{equation}
< \tau_{tun} (0, a) >_{exp} = < t^{ph}_{tun}  >_E
\label{13a}
\end{equation}
because \begin{math}
< t (0) >_{exp} = < t(0) >_{in} .
\end{math}
 
The main criticism, by authors of \cite{land94,lan92}  and also
\cite{leav93,leav95}, of any approach to the definition of tunnelling
times, in which spatial or temporal averaging over moving wave packets
is used, implies the lack of a causal relationship in evolution of an
incoming peak or centroid, turning into an outgoing peak or centroid. It
is clear already from the 60--th (see, for instance, \cite{olkrec69} )
that these reasonings are particularly true only for the spatial
approach with finite (not asymptotic) distances from the interaction
region. And this criticism concerns only the attempts of the authors of
\cite{lan92} to trace evolution of an incoming peak into an outgoing one
but not, strictly speaking, the O--R definition of collision,
tunnelling, transmission, penetration, reflection, durations, etc; our
definition of the mean duration of any such process is, by no means,
based on the assumption that the centroid (or peak) of the incident wave
packet directly evolves into the centroid (or peak) of the transmitted
and reflected packets, as it was erroneously claimed in \cite{leav95},
but does simply signify the difference between the mean time values of
the passing of the final and initial wave packets through the
appropriate points, {\it regardless of any intermediate motion,
transformation and reshaping of these wave packets}. And for any
collision (and so on) processes, as a {\it whole}, one can test the
causality condition. However, there is no unique general formulation of
the causality condition, necessary and sufficient for all possible cases
of collisions (and not only for nonrelativistic wave packets, but also
for relativistic ones). The simplest (or the most strong)
nonrelativistic causality condition implies the non--negative values of
the mean durations. However, this is a {\it sufficient} but {\it not the
necessary} causality condition. Negative times (advance phenomena) were
revealed even near nuclear resonances, distorted by the non--resonant
background (see, in particular, \cite{olk90} ); similarly, advance
phenomena can occur at the beginning of tunnelling (see Sec.6).
Generally speaking, the complete causality condition can be connected
not only with the mean time duration but also with other temporal
properties of the considered process. For example, the following variant
of the causality condition seems to be somewhat more realistic: the
difference between the {\it effective starts} of final and initial
fluxes is non--negative, the effective start being defined as the
difference between the mean instant of wave--packet passing through the
appropriate point and the square root of the corresponding
instant--distribution variance. And also this condition is only
sufficient but not necessary because in many realistic cases wave
packets have infinite and not very rapidly decreasing forward tails.
Much more realistic formulations of the causality conditions for wave
packets with infinite tails are presented in Sec.8.
 
\section{The meaning of the mean dwell time in the light of the
Olkhovsky--recami formalism.}
 
As it is known \cite{jaw} (see also \cite{olkrec94,olkrec95} ) the mean
dwell time can be presented in two equivalent forms:
\begin{equation}
< \tau^{dw} (x_i, x_f ) > =
\left[ \int^{\infty}_{-\infty} {\diff t} \int^{x_2}_{x_1} | \Psi (x, t)
|^2
{\diff x} \right] \left[ \int^{\infty}_{-\infty} J_{in} (x_i, t) {\diff
t} \right]^{-1}
\label{14a}
\end{equation}
and
\begin{equation}
< \tau^{dw} (x_i, x_f ) > =
\left[ \int^{\infty}_{-\infty} t J (x_f, t) {\diff t} -
\int^{\infty}_{-\infty} t J (x_i, t )
{\diff t} \right] \left[ \int^{\infty}_{-\infty} J_{in} (x_i, t) {\diff
t} \right]^{-1}
\label{14b}
\end{equation}
with $-\infty < x_i \leq 0; a \leq x_f < \infty $ . In its primary
definition (\ref{14a}) another, than in Sec.2, measure (weight) was
used in integrating over $t$. What is that measure and what is its
meaning ? Taking into account equation $ \int J_{in} (x_i, t) {\diff t}
= \int | \Psi (x, t) |^2 {\diff x}$ , which follows from the continuity
equation, one can easily see that this measure is $dP(x,t) = | \Psi
(x,t)|^2 dx / \int | \Psi (x,t)|^2 {\diff x}$ and it has the well--known
quantum--mechanical meaning of the probability for a particle to be {\it
found (localized)} or to {\it stay (dwell)}, in the spatial region
$(x,x+ {\diff x})$ at the moment $t$, independently on the motion
processes. Then the quantity $ P(x_1, x_2; t) = \int^{x_2}_{x_1} | \Psi
(x,t)|^2 {\diff x} / \int^{\infty}_{-\infty} | \Psi (x, t)|^2 {\diff x}$
has the evident meaning of the probability of particle dwelling in the
spatial range $(x_i, x_f)$ at the moment $t$ (see also \cite{beg}.
And the equivalency of relations (\ref{14a}) and (\ref{14b}) is a
consequence of the continuity equation which connects the staying
(dwelling) and the motion (traversing, transferring, passing, entering,
outgoing) processes. However,  we note that the applicability of the
measure $P(x_1,x_2;t)$ for the time analysis (in contrast to the space
analysis) is limited since it serves directly for calculations of only
dwelling durations but not of their distributions.
 
Taking into account that $J(x_i, t) = J_{in} (x_i, t) + J_R (x_i, t) +
J_{int} (x_i, t)$ and $J(x_f, t) = J_T (x_f,t) $ with $J_{in}, J_R$ and
$J_T$ corresponding to the wave packets $\Psi_{in} (x_i,t), \Psi_R
(x_i,t)$ and $\Psi_T (x_f,t)$, constructed from the stationary wave
functions $\Psi_{in}, \Psi_R = A_R {\rm exp} (-ikx)$ and $\Psi_T$,
respectively,
\begin{eqnarray*}
J_{int} (x, t) =
{\rm Re} \{ (i \hbar / m) [ ( \Psi_{in} (x, t)) (\partial \Psi_R^*
(x,t)/ \partial x) + ( \Psi_R (x,t) (\partial \Psi^*_{in} (x,t)/
\partial x ) ]  \}
\end{eqnarray*}
and
\begin{eqnarray*}
\int_{-\infty}^{\infty} J_{int} (x_i,t) {\diff t}  =  0,
\end{eqnarray*}
we obtain
\begin{equation}
< \tau^{dw} (x_i,x_f) > = <T >_E < \tau_T (x_i,x_f) > + < R(x_i) >_E <
\tau_R (x_i, x_i) >
\label{15}
\end{equation}
with
\begin{eqnarray*}
<T>_E = < | A_T |^2 v >_E \Bigr/ < v >_E; < R(x_i) >_E = <R>_E +
<r(x_i)>; \\ <R>_E = < | A_R |^2 v >_E \Bigr/ < v >_E, <R>_E + <T>_E = 1
\\ \end{eqnarray*}
and
\begin{eqnarray*}
< r (x ) > =
\int^{\infty}_{-\infty} [  J_+ (x, t)  -
 J_{in} (x, t ) ]
{\diff t}
  \Bigr/  \int_{-\infty}^{\infty} J_{in} (x, t) {\diff
t}.
\end{eqnarray*}
We stress that $< r (x) > $ is negative and tends to $0$ when $x$ tends
to $-\infty$.
 
When $\Psi_{in} (x_i, t)$ and $\Psi_R (x_i, t)$ are well separated in
time, i.e., $< r(x_i) > = 0$, we obtain the simple weighted average rule
\begin{equation}
< \tau^{dw} (x_i, x_f) > = <T>_E <\tau_T (x_i,x_f)> + <R>_E <\tau_R
(x_i, x_i) > .
\label{16}
\end{equation}
For a rectangular barrier with $\chi a \gg 1$ and quasi--monochromatic
particles, the expressions (\ref{15}) and (\ref{16}) with $x_i = 0$ and
$x_f = a$ pass to the known expressions
\begin{equation}
<\tau^{dw} (x_i, x_f) > = < \hbar k /  \chi V_0 >_E
\label{15a}
\end{equation}
(taking account of the interference term $ < r (x_i) > $ ) and
\begin{equation}
<\tau_{dw} (x_i, x_f) > = < 2 /  \chi v >_E
\label{16a}
\end{equation}
(when the interference term $ < r (x_i) > $ is equal to $0$).
When $A_R = 0$, i.e., the barrier is transparent, the mean dwell time
(\ref{14a}), (\ref{14b}) is automatically equal to
\begin{equation}
< \tau^{dw} (x_i, x_f) > = < \tau_T (x_i, x_f) >.
\label{17}
\end{equation}
It is not clear how to define {\it directly} the variance of the
dwell--time distribution.
The approach, proposed in \cite{dum}, is rather sophisticated, with an
artificial abrupt switching on the initial wave packet. It is possible
to define the variance of the dwell--time distribution indirectly, in
particular, by means of relation (\ref{15}), basing on the variances of
the transmission--time and reflection--time distributions, or by means
of relation (\ref{14a}), basing on the variances of the positions $x_1$
and $x_2$.

\section{Analysis of the Larmor and B\"{u}ttiker--Landauer clocks.}
 
One can often realize that introducing additional degrees of freedom as
"clocks" distorts the true value of the tunnelling time. The Larmor
clock uses the phenomenon of changing the spin orientation (The Larmor
precession or spin--flip) in a weak homogeneous magnetic field covered
the barrier region. If initially the particle spin is polarized in the
$x$ direction, after tunnelling the spin develops small $y$ and $z$
components. The Larmor time $\tau^{La}_{y, T}$ and $\tau^{La}_{z, T}$
are
defined by the ratio of the spin--rotation angles around $z$--axis and
$y$--axis (in turn defined by the developed $y-$ and $z$--{\it spin}
components respectively) to the precession (rotation) frequency
\cite{baz,ryb,butt83}. For an opaque rectangular barrier with $\chi a
\gg 1$ the expressions
\begin{equation}
< \tau^{La}_{y, tun} > = < \tau^{dw} (x_i, x_f ) > = < k / \chi V_0 >_E
\label{18}
\end{equation}
and
\begin{equation}
< \tau^{La}_{z, tun} > = < m a / \hbar \chi >_E
\label{19}
\end{equation}
were obtained.
 
In \cite{olkrec92,huang88} it was noted that, if the magnetic field
region is infinite, the expression (\ref{18}) passes into the expression
(\ref{11a}) for the phase tunnelling time, after averaging over the
small energy spread of the wave packet.
 
As to (\ref{19}), in the reality we have not a precession but a jump to
position `` spin--up" or ``spin--down" (spin--flip) accompanyed by the
Zeeman energy--level splitting \cite{gold,pri}. Due to the Zeeman
splitting, the component of the spin, that is parallel to the magnetic
field, corresponds to a higher tunnelling energy and hence it tunnels
preferentially. And, namely, therefore one can realize that this time is
connected with the energy dependance of $ | A_T | $ and coincides with
the expression (\ref{12a}).
 
The work of the B\"{u}ttiker-- Landauer clock is connected with the
modulation cycle (absorption or emission of modulation quanta), caused
by the oscillating part of a barrier, during tunnelling. And also in
this case, one can realize that the coincidence of the
B\"{u}ttiker--Landauer
time with (\ref{12a}) is connected with the energy
dependence of $ | A_T | $ for the same reasons as for $ < \tau^{La}_{z,
tun} >$.
 
\section{Analysis of the mean tunnelling times, defined by averaging
over kinematic paths.}
 
The Feynman path--integral approach to quantum mechanics was applied in
\cite{sok87,pol84,sok90,mug88} to study and calculate the mean
tunnelling time averaged over all paths, that have the same beginning
and end, with the complex weight factor ${\rm exp}[i S (x(t) /  \hbar]$,
where $S$ is the action associated with the path $x(t)$. Namely such
weighting of tunnelling times implies their distribution with a real and
an imaginary components \cite{land94}. In \cite{sok87} the real and
imaginary parts of the obtained complex tunnelling time were found to be
equal to $ \tau^{La}_{y,tun} $ and $- \tau^{La}_{z, tun} $ respectively.
 
An interesting development of this approach, the {\it instanton}
version, is presented in \cite{mug88}. The instanton--bounce path is a
stationary point of the Euclidean action. The latter is obtained by the
analytic continuation to imaginary time in the Feynman path--integrals
containing the factor $ {\rm exp} (iS/ \hbar) $ . This path obeys a
classical equation of motion in the potential barrier with the sign
reversed. In \cite{mug88} the instanton bounces were considered as a
real physical processes. The bounce duration was calculated in real time
and was found to be in good agreement with the one evaluated by the
phase--time method. The temporal density of bounces was estimated in
imaginary time and the obtained result coincided with (\ref{12}) for the
square root of the distribution variance at the limit of the phase--time
approximation. Here one can see a manifestation of the virtual
equivalence of the Schr\"odinger representation and the Feynman
path--integral approach to quantum mechanics.
 
Another definition of the tunnelling time is connected with the Wigner
distribution paths \cite{lee,muga92}. The basic idea of this approach,
finally formulated by Muga, Brouard and Sala, is that the distribution
of the tunnelling times in the dynamical evolution of wave packets
through barriers can be well approximated by a classical ensemble of
particles with a certain distribution function, namely the Wigner
function $f(x,p)$, so that the flux at position $x$ can be separated
into positive and negative components:
\begin{equation}
J(x) = J^+ (x) + J^- (x)
\label{20}
\end{equation}
with
\begin{eqnarray*}
J^+ (x) = \int^{\infty}_0 (p / m) f(x, p) {\diff p}
\end{eqnarray*}
and ${J^- = J - J^+}$ . Then {\it formally} the same expressions
(\ref{3}), (\ref{5}) and (\ref{7}) for the transmission, tunnelling and
penetration durations and so on, as in the O--R formalism, were
obtained with the substitution of $J^{\pm}$ instead of our $J_{\pm}$.
The dwell time decomposition in this approach takes the form
\begin{equation}
<\tau^{dw} (x_i, x_f) > = < T >_E < \tau_T (x_i,x_f) > + < R_M (x_i) >_E
< \tau_R (x_i, x_i) >
\label{21}
\end{equation}
with $ R_M (x) = \int^{\infty}_0 |J^- (x,t) | {\rm d} t$. Asymptotically
$R_M(x)$ tends to $<R>_E$ and (\ref{21}) takes {\it formally} the known
form (\ref{16}).
 
One more alternative is the stochastic method for wave packets
\cite{che90}. It also leads to real times but its numerical
implementation is not trivial \cite{muga92}.
 
In \cite{leav93} the Bohm approach to quantum mechanics was used to
choose a set of classical paths which do not cross. The Bohm formulation
can provide, on the one hand, a strict equivalent to the Schr\"odinger
equation, and on the other hand, a base for the nonstandard
interpretation of quantum mechanics \cite{land94}. The obtained
expression in \cite{leav93} for the mean dwell time is not only positive
definite but gives the unambiguous distinction between particles that
are transmitted or reflected:
\begin{equation}
\tau^{dw} (x_i, x_f) = \int^{\infty}_0 {\diff t} \int^{x_2}_{x_1}
| \Psi (x,t)|^2 {\diff x} = T \tau_T (x_i, x_f) + R \tau_R (x_i, x_i)
\label{22}
\end{equation}
with
\begin{equation}
\tau_T  (x_i, x_f) = \int^{\infty}_0 {\diff t} \int^{x_2}_{x_1}
| \Psi (x,t)|^2 \Theta (x - x_c) {\diff x} / T ,
\label{23}
\end{equation}
\begin{equation}
\tau_R  (x_i, x_f) = \int^{\infty}_0 {\diff t} \int^{x_2}_{x_1}
| \Psi (x,t)|^2 \Theta (x_c - x) {\diff x} / R ,
\label{24}
\end{equation}
where $T$ and $R$ are, here, the mean transmission and reflection
probability respectively, the bifurcation line $x_c = x_c (t)$,
separating transmitted and reflected trajectories, is defined by
relation
\begin{equation}
T = \int^{+\infty}_{-\infty} | \Psi (x,t) |^2 \Theta (x - x_c) {\diff
x}.
\label{25}
\end{equation}
Factually, in addition to the difference in the temporal integration in
this and our formalism $(\int^{\infty}_0$ and
$\int^{\infty}_{-\infty} $ respectively), sometimes essential, this
approach gives one more alternative in separating the flux by the line
$x_c$ :
\begin{equation}
J(x,t) = [ J (x,t)]_T +  [J (x,t)]_R
\label{26}
\end{equation}
with
\begin{eqnarray*}
 [J (x,t)]_T = J (x,t) \Theta [x - x_c (t)] ,
\end{eqnarray*}
\begin{eqnarray*}
[ J (x, t)]_R = J (x,t) \Theta  [ x_c (t) - x ].
\end{eqnarray*}

\section{Peculiarities of the tunnelling evolution.}
 
The results of calculations presented in \cite{olkrec95}, within the
Olkhovsky--Recami formalism, show that: (i) at variance with
\cite{leave93}, no plot for the mean penetration duration of our wave
packets presents any interval with negative values, nor with a
decreasing for increasing $x$; (ii) the mean tunnelling duration does
not depend on the barrier width  $a$ ( the Hartmann--Fletcher effect);
(iii) the quantity $<\tau_{tun} (0,a)>$ decreases when the energy
increases; (iv) the value $<\tau_{pen} (0,x)>$ rapidly increases for
increasing $x$ near $x=0$ and tends to an almost saturation value near
$x=a$.
 
In Fig.2 the dependences  of the values of $< \tau_{tun} (0,a) > $ from
$a$ are presented for electronic wave packets and rectangular barriers
with the same parameters as in \cite{olkrec95} $ (V_0 = 10 {\rm eV};$
mean electron energies $ E=2.5, 5, 7.5 {\rm eV}$ with $\Delta k = 0.02
\AA^{-1}$ (curves 1a, 2a, 3a respectively); energy $E=5 {\rm eV}$ with $
\Delta k = 0.06 \AA^{-1}$ (curves 4a, 5a respectively) ). The curves,
corresponding to different energies and $k$, merge practically into one
curve, 6. And since the dependence of $< \tau^{ph}_{tun} > $ from $a$ is
very weak, the dependance of $ < \tau_{tun} (0,a) > $ from $a$ is
defined  mainly by the dependence of $ < t_+(0) > $ from $a$ (curves
1b--5b). All these calculations manifest the negative value of $< t_+
(0) > $ from $a$ (see also \cite{zai}). Such ``a--causal" advance can be
interpreted as a result of the superposition and interference of
incoming and reflected waves: the reflected--wave packet extinguishes
the back edge of the incoming--wave packet, and the larger is the
barrier width the larger is the part of the back edge of the
incoming--wave packet which is extinguished by the superimposing
reflected--wave packet, up to the saturation, when the contribution of
the reflected--wave packet becomes almost constant, independently from
$a$. Besides all $< t_+ (0) >$ are negative and the values of $<
\tau_{tun} (0,a) >$ are always positive and, moreover, larger than $<
\tau^{ph}_{tun} >$, in accordance with (\ref{13}). In connection with
this , it is relevant to note that the example with a {\it classical}
ensemble of two particles (one with a large above--barrier energy and
the other with a small sub--barrier energy),  presented in \cite{del},
contradicts to our results not only because that tunnelling is a pure
quantum phenomenon without a direct classical limit but, first of
all, because in \cite{del} it is overlooked the fact that the values of
$< t_+ (0) > $ are negative (for our initial condition). The last
calculations of Zaichenko \cite{zai} ( for the same parameters) have
shown that such advance is noticeable also before the barrier front
(however, only near the barrier wall) and, moreover, the values of $<
\tau_{pen} (x_i, x_f) > $ are negative for $x_i = -a/5$ and $x_f = 0,
a/5, 2a/5$ and a little larger values of $x_f$ inside the barrier. But
this result is not a--casual because the causality conditions (see
relations (\ref{39}) and (\ref{39a}) in Sec.8) are fulfilled in this
case.
 
\section{About the analogy between nonrelativistic particle and photon
(electromagnetic wave--packet) tunnelling.}
 
The formal mathematical analogy between the time--dependent quantum
equations for the motion of relativistic particles and the
time--dependent equation for electromagnetic wave propagation was
studied in \cite{gav,ranf95}. Here, we shall deal with the comparison of
the solutions of the time--dependent Schr\"odinger equation for
nonrelativistic particles and of the time--dependent Helmholtz equation
for electromagnetic waves, considering not only the formal mathematical
analogy between them, but also such similarity of the probabilistic
interpretation of the wave function for a particle and of a classical
electromagnetic wave packet, (being according to \cite{akh} the ``wave
function for a single photon") which is sufficient for the identical
definition of mean time instants and durations (and distribution
variances and so on) of propagation, collision, tunnelling,etc.,
processes for particles and photons \cite{agru2}.
 
Concretely, we consider a hollow narrowed rectangular waveguide like
depicted in Fig.3 (with cross section $a \times  b$ of the narrow part,
$a < b$
), which was employed for the experiments with microwaves \cite{end92}.
Inside it, the time--dependent wave equation for any of vector
quantities $\overrightarrow{A}, \overrightarrow{E}, \overrightarrow{H},
( \overrightarrow{A} $ is the vector potential with the subsidiary gauge
condition $div \overrightarrow{A} = 0; \overrightarrow{E} = -(1/c)
\partial \overrightarrow{A} / \partial t$  is the electric field
strength; $ \overrightarrow{H} = rot \overrightarrow{A} $ is the
magnetic field strength ) is
\begin{equation}
\Delta \overrightarrow{A} - (1/c^2) \partial^2 \overrightarrow{A}
/ \partial t^2 = 0.
\label{27}
\end{equation}
As it is known, (see, for instance \cite{jac,mor,bri}), for boundary
conditions %
\begin{eqnarray}
E_y = 0 \;\;\; &{\rm for}& \; z = 0, \;\; {\rm and} \; z = a,
\nonumber
\\
\label{28}
\\
E_z = 0 \;\;\; &{\rm for}& \; y = 0, \;\; {\rm and} \; y = b,
\nonumber
\end{eqnarray}
the monochromatic solution of (\ref{27}) can be represented as a
superposition of following waves:
\begin{eqnarray}
E_x \;\; &=& 0,
\nonumber
\\
E^{\pm}_y \;\; &=& E_0 {\rm sin}(k_z z) {\rm cos} (k_y y) {\rm exp}[i (
\omega t \pm \gamma x ) ],
\label{29}
\\
E^{\pm}_z \;\; &=& -E_0 (k_y / k_z) {\rm sin}(k_y y) {\rm cos} (k_z z)
{\rm exp}[i ( \omega t \pm \gamma x ) ],
\nonumber
\end{eqnarray}
(we have chosen for definiteness $TE$--waves) with $k^2_z + k^2_y +
\gamma^2 = \omega^2/c^2 = (2 \pi / \lambda )^2 ; \; k_z = m \pi /a; k_y
= n \pi / b; m $ and $n$ being integer numbers. So,
\begin{eqnarray}
\gamma \;\; &=& 2 \pi [ ( 1/ \lambda )^2 - (1/ \lambda_c )^2 ]^{1/2},
\nonumber
\\
\label{30}
\\
(1/ \lambda_c )^2 \;\; &=& (m/2a)^2 + (n/2b)^2,
\nonumber
\end{eqnarray}
where $\gamma$ is real $( \gamma = {\rm Re} \gamma )$ if
$\lambda < \lambda_c$ and $ \gamma $ is imaginary $ (\gamma = i
\chi_{em})$ if $ \lambda > \lambda_c $.
Similar expressions for $\gamma$ were obtained for $TH$--waves
\cite{end92,mor}.
 
Generally a solution of (\ref{27}) can be written as a wave packet
constructed from monochromatic solutions (\ref{29}), similarly to a
solution of the time--dependent Schr\"odinger equation for
nonrelativistic particles in the form of a wave packet constructed from
monochromatic terms. Moreover, in the primary--quantization
representation, a probabilistic single--photon wave function is usually
described by a wave packet for $\overrightarrow{A} $ \cite{akh,sch},
for example
\begin{equation}
\overrightarrow{\mathstrut A} (\overrightarrow{\mathstrut r}, t )  =
\int\limits_{k_0 > 0} \displayfrac{\diff^{\: 3} \,
\overrightarrow{\mathstrut k}}{k_0} \overrightarrow{\mathstrut \kappa}
(\overrightarrow{\mathstrut k} ) {\rm exp}
(i\overrightarrow{\mathstrut k}\overrightarrow{\mathstrut r} - i k_0 t
) \label{31}
\end{equation}
in the case of the plane waves, where $ \overrightarrow{r} = \{x, y,
z\};
\overrightarrow{\mathstrut \kappa} (\overrightarrow{\mathstrut k}) =
\sum^2_{i=1}
\kappa_i (\overrightarrow{\mathstrut k}) \overrightarrow{\mathstrut e}_i
(\overrightarrow{\mathstrut k});
\overrightarrow{\mathstrut e}_i \overrightarrow{\mathstrut e}_j =
\delta_{ij};
\overrightarrow{\mathstrut e}_i (\overrightarrow{\mathstrut k})
\overrightarrow{\mathstrut k} = 0,
i,j = 1,2 $ (or $y,z $ if $\overrightarrow{\mathstrut k}
\overrightarrow{\mathstrut r} =
k_x x); k_0 = \omega/c =\epsilon / \hbar c; k = |
\overrightarrow{\mathstrut k}
| = k_0, $ and $\kappa_i ( \overrightarrow{\mathstrut k})$ is the
amplitude for the photon to have momentum $\overrightarrow{k}$ and
$i$--polarization , and $ | \kappa_i
(\overrightarrow{k}) |^2 {\diff^{\: 3} \, \overrightarrow{\mathstrut
k}}$ is then proportional to the
probability that the photon has a momentum between $\overrightarrow{k}$
and $\overrightarrow{k} + {\diff \overrightarrow{k}} $
in the polarization state $\overrightarrow{e}_i$. Though it is not
possible to localize photon in the direction of its polarization,
nevertheless, in a
certain sense, for the one--dimensional propagation, it is possible to
use the space--time probabilistic interpretation of (\ref{31}) along
$x$--axis (the propagation direction) \cite{olk83}.  Usually one uses
not the probability density and the probability flux density with the
corresponding continuity equation {\it directly} but the {\it energy
density} $s_0$ and the {\it energy flux density} $s_x$ (although, in
general, they represent components of not a 4--dimensional vector but
the energy--momentum tensor) with the corresponding continuity equation
\cite{akh} which we write in the two--dimensional ({\it spatially},
one--dimensional) form:
\begin{equation}
\partial s_0 / \partial t + \partial s_x / \partial x  = 0
\label{32}
\end{equation}
where
\begin{equation}
s_0 = [ \overrightarrow{E^*} \;\; \overrightarrow{E}  +
\overrightarrow{H^*} \;\;
\overrightarrow{H} ] / 4 \pi,  \;\;s_x = c [ \overrightarrow{E^*} \;\;
\overrightarrow{H} ]_x / 8 \pi
\label{33}
\end{equation}
and $x$--axis is directed along the motion direction (the mean momentum)
of the wave packet (\ref{31}). We stress that for the spatially
one--dimensional propagation the energy--momentum tensor of the
electromagnetic field reduces to the two--component quantity, scalar
term $s_0$ and 1--dimensional vector term $s_x$, for which continuity
equation (\ref{32}) is {\it Lorentz--invariant}. Then, as a
normalization condition, one choose the equality of the spatial
integrals
of $s_0$ and $s_x$  to the mean photon energy and the mean photon
momentum respectively or simply the unit energy flux density $s_x$. With
this, by passing the problem of the impossibility of the {\underbar
{direct}} space probabilistic interpretation of (\ref{31}), we can
define {\it conventionally} the probability density
\begin{equation}
\rho_{em} \diff x = S_0 \diff x \;\; \Bigr/ \int S_0 \diff x, \;\;  S_0
= \int s_0 \diff y \diff z
\label{34}
\end{equation}
of a photon to be found (localized) in the spatial interval $(x, x +
\diff x)$ along $x$--axis at the moment $t$, and the flux probability
\begin{equation}
J_{em} \diff t  =  S_x \diff t \;\; \Bigr/ \int S_x \diff t, \;\; S_x =
\int s_x \diff y \diff z
\label{35}
\end{equation}
of a photon to pass through point (plane) $x$ in the time interval $(t,
t + \diff t)$, quite similarly to the probabilistic quantities for
particles. The justification and convenience of such definitions are
also supported by the coincidence of the wave--packet group velocity and
the velocity of the energy transport which was established for
electromagnetic waves (at least, in the case of usual plane waves) in
\cite{fel}. Hence, (i) in a certain sense, for time analysis along the
motion direction, the wave packet (\ref{31}) is quite similar to a wave
packet for nonrelativistic particles and, (ii)  similarly to the
conventional nonrelativistic quantum mechanics, one can define the mean
time of photon (electromagnetic wave packet) passing through point $x$
\cite{olk83} :
\begin{equation}
< t(x)> = \int^{\infty}_{-\infty} t J_{em,x} \diff t =
\int^{\infty}_{-\infty} t S_x (x,t) \diff t  \;\; \Bigr/
\int^{\infty}_{-\infty} S_x (x,t) \diff t
\label{36}
\end{equation}
(where for the natural boundary conditions,  $\kappa_i(0) = \kappa_i
(\infty) = 0 $, in energy representation $(\epsilon = \hbar c k_0 )$,
we can use the same form of time operator as for particles in
nonrelativistic quantum mechanics  and hence verify the
equivalence of calculations of $<t(x)>, Dt(x) $, etc.,  in both time and
energy representations). Then, one can use the same interpretation for
the propagation of electromagnetic wave packets (photons) in media and
waveguides when collisions, reflections and tunnelling can take place.
In particular, for waveguides ,like depicted in Fig.3, with boundary
conditions (\ref{28}) and, decreasing and increasing waves when $k_x =
\gamma = i \chi_{em}$.
 
In the case of fluxes  which change their signs with time, we introduce,
following \cite{olkrec92,olkrec94,olkrec95}, quantities $ J_{em,x,\pm}
 = J_{em,x} \Theta (\pm J_{em,x} ) $ with the same physical meaning as
for particles. Therefore, expressions for mean values and variances of
distributions of propagation, tunnelling, transmission, penetration, and
reflection durations can be obtained in the same way as in the case of
nonrelativistic quantum mechanics for particles (with the substitution
of $J$ by $J_{em}$). In the particular case of quasi--monochromatic wave
packets, using the stationary--phase method under the same boundary and
measurement conditions, as considered in Sec.2 for particles, we obtain
the identical expression for the {\it phase tunnelling time}
\begin{equation}
\tau^{ph}_{tun,em} = 2 / c \chi_{em} \;\; {\rm for} \;\; \chi_{em} L \gg
1.
\label{37}
\end{equation}
From (\ref{37}), we can see, that when $ \chi_{em} L > 2 $ the effective
tunnelling velocity
\begin{equation}
v^{eff}_{tun} = L /  \tau^{ph}_{tun,em}
\label{38}
\end{equation}
is more than $c$, i.e., {\it superluminal}. This result agrees with the
results of the microwave--tunnelling measurements presented in
\cite{end92} (see also \cite{ste93}, where, moreover, the effective
tunnelling velocity was identified with the group velocity of the final
wave packet corresponding to a single photon).
 
\section{A remark on reshaping (reconstructing) phenomenon.}
 
The superluminal phenomena, observed in the experiments with tunnelling
photons and evanescent electromagnetic waves
\cite{end92,ste93,ran93,spi94}, generated a lot of discussions on
relativistic causality. And, in connection with this, also an interest
for similar phenomena, observed for the electromagnetic pulse
propagation in a dispersive medium \cite{gar,cri,chu82}, was revived.
Already for long there was ascertained that the {\it wave--front
velocity} of the electromagnetic pulse propagation, when pulses have a
step--function envelope, cannot exceed the velocity of light $c$ in
vacuum \cite{bri,som}. There, the {\it signal velocity} was also
defined as the velocity of the propagation of the pulse main part in
a medium which was shown to be less than the wave--front velocity. These
conclusions were confirmed
by various methods and in various processes, including tunnelling
\cite{ranf95,fox,deu,hass94,azb,hei94}. In \cite{bri} the distinctions
between the above mentioned velocities and the {\it group velocity}
were also analyzed.
 
One of the argued problems consists in the absence of a step--function
form of forward edges for realistic wave packets \cite{ranf95,hei94}.
In such cases the conclusions of \cite{bri} can seem to be inapplicable.
Nevertheless an infinite but sufficiently rapidly decreasing forward
edge of a pulse can be cut off, {\it with any desired degree of the
accuracy}, (defined, for instance, by a sensitivity of registration
devices or a chosen mathematical approximation), without an essential
distortion of the pulse spectral expansion. This can give a possibility
to apply the conclusions of \cite{bri,som}. But independently from these
reasonings, one can search a principal understanding of cases with
superluminal group velocities ( or effective velocities, like
(\ref{38}))
{\it without} violations of special relativity or causality.
 
A possible way of such understanding can consist in explaining the
superluminal phenomena during tunnelling on the base of a pulse {\it
attenuated reshaping (or reconstructing)} discussed at the classical
limit earlier by \cite{gar,cri,chu82}. The later parts of an input pulse
are preferentially attenuated in such a way that the output peak appears
shifted toward earlier times, arising from the forward tail of the
incident pulse in a {\it strictly causal manner} \cite{ste93}. In
particular, the following reasonable scheme
is quite compatible with the usual idea of causality: if an overall
pulse attenuation is very strong and, in the same case, during
tunnelling, the leading edge of the pulse is less attenuated than the
trailing edge, then, the time envelope of the out--coming final small
flux
can be totally {\it under} the temporal initial--flux envelope, which
should pass through the same position if its motion were free in vacuum
(see also the discussion in \cite{nimt94,stei94,nimtz94}). And, if the
dependence of $A_T$ from energy is much more weak than the dependence of
the weight factor in an initial wave packet, the spectral expansion and
hence the geometrical form of the transmitted  wave packet will be
practically undistorted in comparison with the spectral expansion and
the form of
the initial wave packet (reshaping). But if the dependence of $A_T$ from
energy is not weak, then the pulse form and width can be noticeably
changed (reconstruction).
 
The proposed scheme can be considered as a possible {\it sufficient}
(but {\it not necessary}) causality condition and one can try to
formulate more general causality condition when the temporal envelope of
the final flux can even go out of the temporal envelope of the initial
pulse flowing through the same position. Really, one can assume that the
wave--packet spectral expansion and then the shape and width of the
final
pulse remain the same as for the initial pulse (reshaping). And, if the
dependence of $A_T$ from energy is not weak, then the pulse form and
width can be changed (reconstructing). For example, the following
relation
\begin{equation}
\int^T_{-\infty} [ J_{in} (x_f,t) - J_{f,+} (x_f,t) \diff t \ge 0,
  \;\; -\infty < T < \infty
\label{39}
\end{equation}
is quite acceptable. It does simply signify that during any
semi--confined
(from above) time interval, an integral final flux (along any direction)
does not exceed that integral flux which should pass through the same
position during the free motion (with the light velocity $c$ for
photons) of the initial wave packet in vacuum, although, by the way, one
can find such finite $T_1$ and $T_2$ , $(-\infty < T_1 < T_2 < \infty)$
for which
\begin{eqnarray*}
\int^{T_2}_{T_1} [ J_{in} (x_f,t) - J_{f,+} (x_f,t) ] \diff t < 0.
\end{eqnarray*}
One can also propose another causality condition:
\begin{eqnarray}
\int^{T_0}_{-\infty} &t& J_{f,+} (x_f,t) \diff t \;\;  \Bigr/
\int^{T_0}_{-\infty} J_{f,+} (x_f,t) \diff t\nonumber
\\
\label{39a}
\\
&-& \int^{T_0}_{-\infty} t J_{in} (x_f,t) \diff t \;\;  \Bigr/
\int^{T_0}_{-\infty} J_{in} (x_f,t) \diff t  \ge 0\nonumber
\end{eqnarray}
where $T_0$ is the instant of the intersection of the temporal envelopes
of both fluxes {\it after} the final--peak appearance. Relation
(\ref{39a}) does simply signify that there is a {\it delay} in the
averaged appearance of the forward part of the final wave packet, in
comparison with the averaged appearance of the forward part of that wave
packet, which should pass through the same position $x_f$, during the
free
motion of the initial wave packet in vacuum. The conditions (\ref{39})
and (\ref{39a}) are much more general than the previous one. The same
relations can be also used for the nonrelativistic causality conditions,
with the only substitution of $x_f$ by $x_i$ in $J_{in}$.
 
It is curious that, without violating such causality, a certain
undistorted {\it information}, carried out by a low--frequency
modulation of a high--frequency wave packet, can be transmitted
(however, with a strong attenuation) with a
{\it superluminal} wave--packet group (or effective) velocity when
attenuated reshaping takes place.
 
\section{Conclusions and prospects}
 
I. Now one can conclude that the basic quantum--physical formalism for
determining the collision and tunnelling times for nonrelativistic
particles and for photons has been already, at least in principle,
constructed: (1) there are self--consistent definitions of mean
time instants and time durations of various collision processes
(including tunnelling) together with variances of their distributions,
based on utilizing the properties of time as a {\it quantum--physical
observable} (in quantum mechanics and in quantum electrodynamics), just
similarly to other observables; (2) these definitions are functioning
rather well, at least for {\it asymptotic} distances between initial
wave
packets and interaction regions and {\it finite} distances between final
wave packets and interaction regions. In these cases, the phase--time,
clock and instanton approaches give the results which are coincident
with the mean duration or the square root of the duration--distribution
variance obtained within our formalism. And, the asymptotic mean dwell
time is the weighted average sum of the corresponding tunnelling and
reflection durations. Moreover, such `` asymptotic" coincidence can
be naturally extended, if we take into account, also, the mean squared
time duration
\begin{equation}
< [ \tau_N (x_i, x_f)^2 ] >  =  [ < \tau_N (x_i, x_f) > ]^2 + D \tau_N
(x_i, x_f)
\label{40}
\end{equation}
with
\begin{eqnarray*}
D \tau_N (x_i, x_f) = D t_n (x_f) + D t_+ (x_i)
\end{eqnarray*}
where index $N$ signifies $T$ or $tun$ or $pen$ or $R$, and $n = + $ or
$-$ . Relation (\ref{40}) can be rewritten also in the following
equivalent forms:
\begin{eqnarray}
< [ \tau_N (x_i, x_f)]^2 >  =  < [ t_n (x_f) - < t_+ (x_i) > ]^2  > + D
t_+ (x_i)\nonumber
\\
\label{40a}
\\
 = < [ < t_n (x_f) > - t_+ (x_i) ]^2 > + D t_n (x_i).\nonumber
\end{eqnarray}
And, now we see that $ [ < \tau^{ph}_T > ]^2 + D \tau^{ph}_T $ , the
squared {\it hybrid time} $ [ ( \tau^{La}_{y,tun} )^2 + (
\tau^{La}_{z,tun} )^2 ]^{1/2}$ , introduced by B\"{u}ttiker
\cite{butt83}, and the squared absolute value
of the complex tunnelling time, in the Feynman path--integration
approach,
are examples of mean squared durations and, all three are coincident in
the cases of the infinite spatial extension of the magnetic field for
the B\"{u}ttiker hybrid time and of the instanton version pf the
Feynman formalism.
By the way, this formalism has been earlier applied and tested in the
time analysis of nuclear and atomic collisions for which the boundary
conditions are experimentally and theoretically assigned in the region,
asymptotically distant from the interaction region, where the incident
(before collision) and final (after collision) fluxes are well separated
{\it in time} , without any superposition and interference. And, it has
been supported  (see in particular, \cite{olk84,olk90}
and references therein)  by the results: (i) the validity of a
correspondence
principle between the time--energy QM commutation relation and the CM
Poisson brackets; (ii) the validity of an Ehrenfest principle for the
average time durations; (iii) the coincidence of the quasi classical
limit of our own QM definitions for time durations (when such a limit
exists; i.e., for above--barrier energies) with analogous well-known
expressions of classical mechanics; (iv) the analysis of all other
known theoretical approaches to the definition of collision durations,
on the base of our formalism; (v) the analysis of experimental data on
direct and indirect measurements of nuclear--reaction durations, at the
range $10^{-21}$--$10^{-15}$ sec, and, in particular, the extraction
from
these data of informations on compound--nucleus level densities with the
appropriate juxtaposition of the obtained informations with data of
other experiments.
 
Let us stress that for complete extracting the time--duration values
from experimental data on indirect measurements of nuclear--reaction
durations, it is necessary to utilize not only the expressions for mean
durations but also correct definitions of the duration variances and the
higher--order central moments of the duration distribution \cite{olk90}
which is
provided by our formalism. At least, let us note that such a formalism
also provided useful tools for resolving some long-- standing problems
related to the {\it time--energy uncertainty relation}
\cite{olk84,olk90}.
 
II.  For the applications of the presented formalism to such cases, when
one intends to consider, not only asymptotic distances but also the
region {\it inside and near} interactions, we have revised the notation
of averaging weight (or integration measure) in time representation,
utilizing two measure $J_{\pm} (x,t)$ for calculations of instant and
duration mean values, distribution variances, for particle moving,
passing, transferring and the measure $\diff P (x,t)$ or $ P
(x_1,x_2;t)$
for limited calculations of only mean durations for particle staying, or
dwelling. And, with these three measures, we can arrange all known
approaches (but, of course, within the {\it conventional quantum
mechanics}), including the mean dwell time, the Larmor--clock times and
the times given by various versions of the Feynman path--integration
approach, into an unique {\it consistent} and {\it non--contradictory
scheme} on the base of our formalism {\underbar{even}} inside and near
barriers.
 
By the way, $\diff P (x,t)$ can be used as the adequate averaging
measure for defining mean values and variances of position and distance
distributions (see also \cite{olkrec69}), together with $J_{\pm} (x,t)$
as the measures for the definitions of the mean traversed distances
during finite time intervals, for constructing the basic
quantum--mechanical formalism of the  space analysis of collision and
propagation processes.
 
III. The O--R flux separation scheme, within the  conventional
quantum mechanics (and quantum electrodynamics), is not the only
possible
one,although it is the only known {\it incoherent} flux separation
without introducing any new postulates. Within the conventional quantum
theory one can also get the physically clear (but mathematically not
very suitable) {\it coherent} wave--packet separation by positive and
negative momenta, which is explicit out of the barrier region and is
obtainable by the momentum Fourier expansion inside the barrier region.
This separation can be transformed in the incoherent flux separation,
after utilizing the postulate of {\it quantum measurement theory},
about a possibility to describe measurement conditions by the
corresponding projector, acting on wave functions, namely by the
projectors $\Lambda_{exp,\pm}$ onto positive--momentum and
negative--momentum states respectively. There are also flux separation
schemes within nonstandard versions of quantum theory (see examples in
Sec.5). However, whatever separation scheme we choose, we have to keep
to at least two necessary conditions: (1) the probabilistic meaning of
every normalized flux component and, (2) passing to the standard flux
expressions in the asymptotically remote spatial region, well--known in
quantum collision theory (since the boundary conditions of {\it any}
quantum collision in the asymptotic range are for long inspected and
have not to depend from a chosen version of quantum mechanics).
 
For the space inside and near a barrier at least four kinds of
separations of wave--packet fluxes do now coexist, with the fulfillment
of these conditions for
all of them. These separations are defined by different schemes,
although univocally from the mathematical point of view:
(i) the O--R separation $ J = J_+ + J_- $ with $ J_{\pm} = J \Theta (
\pm J ) $ was obtained within the conventional probabilistic  continuity
equation (following from the time--dependent Schr\"odinger equation)
without any new physical postulate or any new mathematical approximation
\cite{olkrec95}. The asymptotic behaviour of the obtained expressions
was tested by the comparison with other approaches and with the
experimental results \cite{olkrec92};
(ii) the proposed here separation $ J = J_{exp,+} + J_{exp,-}
(J_{exp,\pm} $ being the fluxes which correspond to $
\Lambda_{exp,\pm} \Psi (x,t) $ respectively) obtained also within the
conventional probabilistic continuity equation, however after applying
the projector ( or wave--function {\it reduction}) postulate of quantum
measurement theory. The asymptotic behaviour of the expressions,
obtained on the base of this separation, is the same as in $(i)$;
(iii) relation (\ref{20}) was obtained in the Muga--Brouard--Sala
approach, according to the physical clear incoherent flux separation by
positive and negative momenta, but with additional introducing the model
of the Wigner--path distribution;
(iv) relation (\ref{26}) was obtained in the Leavens approach,
according to the incoherent flux separation by trajectories of particles
to be transmitted and to be reflected, with introducing the nonstandard
Bohm interpretation of quantum mechanics.
 
The flux separation schemes (i), (iii), and (iv) give {\it asymmetric}
expressions for the mean dwell time near a barrier (see the relations
(\ref{15}), (\ref{21}), and (\ref{22})--(\ref{25}) respectively),
apparently due to the right--left asymmetry of boundary conditions: we
have a partially simultaneous coexistence of incident and reflected wave
packets from the left and the only one transmitted wave packet on the
right. The separation (ii) gives the {\it symmetric} expression
(\ref{16}) for the mean dwell time even near a barrier.
 
IV. From reasonings, presented in Sec.2, 6, and 8, one can easily
see that positive values (or values inside (or on) the light cone for
relativistic waves) of the collision or propagation or tunnelling
duration is only a sufficient but not the necessary causality condition.
Now, we have not a unique general formulation of the causality principle
which would be necessary for all possible cases. In Sec.2 and Sec.8 some
new formulations of the causality condition are proposed for possible
approbations.
 
V. The phenomenon of reshaping (reconstructing), which was spoken about
in Sec.8, as well as the advance phenomenon at the beginning of
tunnelling, which was spoken about in Se.6, are closely connected with a
coherent super--position of incoming and reflected waves. It is
advisable
to examine these phenomena from various viewpoints and within various
approaches with the scope to elucidate {\it causality} condition during
tunnelling. Moreover, the investigation of the phenomena of reshaping
(reconstructing) and advance by themselves could pursue, moreover, two
important purposes: it will inevitably be the necessary part of the
future {\it kinematic} theory of the tunnelling of particles, waves,
many--particle systems, and solitons inside and near potential barriers
and besides, that can serve as a base of the birth and development of a
new field of the physical information--{\it a superluminal}
propagation of information (see also \cite{hei94}).
 
VI. It is known that there are multiple internal reflections from the
both potential walls and corresponding multiple penetrations through the
walls during the particle motion inside a potential well or a potential
barrier with above--barrier energies \cite{mcv,and89}. The sums of the
multiply reflected and penetrated waves give the resulted reflected and
transmitted waves with the final reflection and transmission
amplitudes. Naturally, the following question arises: are there such
multiple internal reflections and corresponding penetrations during
particle tunnelling with sub--barrier energies ? In \cite{and89} this
question was studied and replied formally positively. But a simple
analysis permits to clear up that the matching conditions give definite
solutions only when inside a barrier the motion along the incident flux
is described by the decreasing wave and the motion against the incident
flux is described by the increasing wave. However, for such waves, the
fluxes are always equal to zero. For resolving this paradox, one can try
to analyze the momentum Fourier--expansions of decreasing and increasing
waves. But then intricate paradoxes with a-causality appear. This
problem is very curious and can born some surprises.
 
VII. As one can conclude, the O--R formalism, presented here, permits in
principle to study temporal characteristics in the Schr\"odinger and
Feynman representations (which are formally equivalent). By the way, an
interesting attempt was undertaken in \cite{ian} to develop the
self--consistent method for calculating time quantities related to the
motion of a particle, utilizing the Feynman representation and comparing
the proposed method with the O--R formalism (in its earlier version
presented in \cite{olkrec92}, however without the separation $ J = J_+ +
J_- )$.
 
VIII. There is one more possible formally equivalent to the
Schr\"odinger
and Feynman representation for examining the collision and tunnelling
evolution. As it is known, in the quantum theory to energy $E$, two
operators correspond-- the operator $ i \hbar \partial / \partial t $
and the hamiltonian operator ${\cal H} $ in terms of the coordinate and
momentum operators. The duality of these operators is well seen from the
Schr\"odinger equation $ {\cal H} \Psi = i \hbar \partial \Psi /
\partial
t $ . The similar duality takes place for time in quantum mechanics:
besides the general form $ -i \hbar \partial / \partial E$, which is
valid for any physical system (in the continuum energy spectrum), it is
possible to express the time operator ${\cal T}$, utilizing the
commutation relation $ [{\cal T , H } ] = i \hbar $, in terms of the
coordinate and momentum operators too \cite{ros,olk90}. And, to study
the
collision and the tunnelling evolution, {\it via} the operator ${\cal
T}$
with the corresponding equation $ {\cal T} \Psi = t \Psi $, it
can prove to be useful too, particularly for researching, the influence
of the barrier form on the tunnelling time \cite{olk90}.
 
IX. Time analysis of more complex processes, such as formations and
decays of metastable states and time correlations of fluctuations of
various quantities in many--particle systems with the accompaniment of
tunnelling processes, has to be developed on the base of the adequate
formalisms for such processes, which are not developed yet and were
only marked in \cite{olk90} for simple approximations.
 
X. Time analysis in processes in the discrete energy spectra (for
instance, for evolving wave packets composed from bound states inside
two--well potentials with a barrier between wells) is quite different
from time analysis of processes in the continuous energy spectra. For
such processes, one can use the formalism based on the properties of the
time operator in the discrete energy spectra \cite{olk90,agru2}, and,
 durations of transitions, (with non--zero fluxes), from one well to
another, are defined by the Poincare` period $ 2\pi \hbar / d_{min} $,
where $ d_{min} $ is the maximal common divisor of the level distances.
The latter is defined, mainly, by the minimal level splitting caused by
the barrier and hence depends on the barrier transition (penetration)
probability at the appropriate energies.
 
\par

\par
\newpage
 
\begin{center}
FIGURE CAPTIONS
\end{center}
 
\vspace{1 cm}
 
\begin{description}
 
\item Fig. 1 - The incoming, reflected and transmitted plane waves in
the stationary picture of a particle, colliding with a barrier and
tunnelling through it.
 
\item Fig. 2 - Plots of $ < t_+ (a) > $ and $ < t_+ (0) > $ as functions
of the variable $a$  for different values of $E$ and $\Delta k $ of
Gaussian wave packets.
 
\item Fig. 3 - The rectangular waveguide with narrow--part section of
dimension $ b $ and length $ L $.
 
\end{description}

\end{document}